\documentstyle{cupconf}

\def\fnum@figure{{\rm Fig.\space\thefigure.---}}
\def\fnum@table{{\rm Table \thetable:}}
\def\fps@figure{bp}
\def\fps@table{bp}
\def\eps@scaling{.95}
\def\epsscale#1{\gdef\eps@scaling{#1}}
\def\plotone#1{\centering \leavevmode \epsfxsize=\eps@scaling\columnwidth \epsfbox{#1}}

\def\plotfiddle#1#2#3#4#5#6#7{\centering \leavevmode \vbox to#2{\rule{0pt}{#2}}
\includegraphics{#1}}

\def \AAL #1 #2 {{\em Astron. Astrophys. Lett.\/} {\bf #1}, L#2}
\def \AAP #1 #2 {{\em Astron. Astrophys.\/} {\bf #1}, #2}
\def \AAR #1 #2 {{\em Astron. Astrophys. Rev.\/} {\bf #1}, #2}
\def \AAS #1 #2 {{\em Astron. Astrophys. Suppl. Ser.\/} {\bf #1}, #2}
\def \AJ #1 #2 {{\em Astron. J.\/} {\bf #1}, #2}
\def \ANNREV #1 #2 {{\em Ann. Rev. Astron. Astrophys.\/} {\bf #1}, #2}
\def \APJ #1 #2 {{\em Astrophys. J.\/} {\bf #1}, #2}
\def \APJL #1 #2 {{\em Astrophys. J. Lett.\/} {\bf #1}, L#2}
\def \APJS #1 #2 {{\em Astrophys. J. Suppl.\/} {\bf #1}, #2}
\def \APSS #1 #2 {{\em Astrophys. Space Sci.\/} {\bf #1}, #2}
\def \IAUC #1    {{\em IAUC\/} {#1}}
\def \MN #1 #2 {{\em Mon. Not. R. Astr. Soc.\/} {\bf #1}, #2}
\def \MEM #1 #2 {{\em Mem. R. Astr. Soc.\/} {\bf #1}, #2}
\def \MESS #1 #2 {{\em The Messenger\/} {\bf #1}, #2}
\def \NAT #1 #2 {{\em Nature\/} {\bf #1}, #2}
\def \PLR #1 #2 {{\em Phys. Lett. Rev.\/} {\bf #1}, #2}
\def \PASJ #1 #2 {{\em Publ. Astron. Soc. Japan\/} {\bf #1}, #2}
\def \PASP #1 #2 {{\em Publ. Astr. Soc. Pacific\/} {\bf #1}, #2}
\def \SAIT #1 #2 {{\em Mem.\ Soc.\ Astron.\ It.\/} {\bf #1}, #2}

\newcommand{\kms}{km~s$^{-1}$}
\newcommand{\ie}{{\it i.e., }}
\newcommand{\etal}{{\it et al. }}
\newcommand{\eg}{{\it e.g., }}
\newcommand{\hst}{{\it HST}}

\def\lesssim{\mathrel{\hbox{\rlap{\hbox{\lower4pt\hbox{$\sim$}}}\hbox{$<$}}}}
\def\gtrsim{\mathrel{\hbox{\rlap{\hbox{\lower4pt\hbox{$\sim$}}}\hbox{$>$}}}}

\def\arcdeg{\hbox{$^\circ$}}
\def\arcmin{\hbox{$^\prime$}}

\def\fs{\hbox{$.\!\!^{\rm s}$}}

\def\farcs{\hbox{$.\!\!^{\prime\prime}$}}

\begin{document}

%-----title and author----------------------

\title[Radio Supernovae and GRB~980425]{Radio Supernovae and GRB~980425} 

\author[K. W. Weiler {\it et al.\/}]%
{K\ls U\ls R\ls T\ns W.\ns W\ls E\ls I\ls L\ls E\ls R$^1$,\ns N\ls I\ls N\ls O\ns P\ls A\ls N\ls A\ls G\ls I\ls A$^2$\thanks{On assignment from the Astrophysics Division, Space Science Department of ESA.},\ns\\
R\ls I\ls C\ls H\ls A\ls R\ls D\ns A.\ns S\ls R\ls A\ls M\ls E\ls K$^3$,\ns
S\ls C\ls H\ls U\ls Y\ls L\ls E\ls R\ns D.\ns V\ls A\ls N\ns D\ls Y\ls K$^4$,\ns\\ 
M\ls A\ls R\ls C\ls O\ls S\ns J.\ns M\ls O\ls N\ls T\ls E\ls S$^5$,\ns 
\and\ns 
C\ls H\ls R\ls I\ls S\ls T\ls I\ls N\ls A\ns K.\ns L\ls A\ls C\ls
E\ls Y$^6$\thanks{NRC Postdoctoral Fellow.}}

\affiliation{$^1$NRL, Code 7213, Washington, DC 20375-5320; weiler@rsd.nrl.navy.mil\\[\affilskip]
$^2$Space Telescope Science Institute, 3700 San Martin Drive, Baltimore, MD 21218; panagia@stsci.edu\\[\affilskip]
$^3$P.O.~Box 0, NRAO, Socorro, NM 87801; dsramek@nrao.edu\\[\affilskip]
$^4$IPAC/Caltech, Mail Code 100-22, Pasadena, CA 91125; vandyk@ipac.caltech.edu\\[\affilskip]
$^5$NRL, Code 7212, Washington, DC 20375-5320; montes@rsd.nrl.navy.mil\\[\affilskip]
$^6$NRL, Code 7213, Washington, DC 20375-5320; lacey@rsd.nrl.navy.mil}
\setcounter{page}{1}

\maketitle

%-------------------------------------------------------

\begin{abstract}
Study of radio supernovae (RSNe) over the past 20 years includes two
dozen detected objects and more than 100 upper limits. From this work
we are able to identify classes of radio properties, demonstrate
conformance to and deviations from existing models, estimate the
density and structure of the circumstellar material and, by inference,
the evolution of the presupernova stellar wind, and reveal the last
stages of stellar evolution before explosion.  It is also possible to
detect ionized hydrogen along the line of sight, to demonstrate binary
properties of the stellar system, and to show clumpiness of the
circumstellar material. More speculatively, it may be possible to
provide distance estimates to radio supernovae.

The interesting and unusual radio supernova SN 1998bw, which is
thought to be related to the $\gamma$-ray burst GRB~980425, is
discussed in particular detail.  Its radio properties are compared and
contrasted with those of other known RSNe.
\end{abstract}

\section{Introduction}

A series of papers published over the past 20 years on radio
supernovae (RSNe) has established the radio detection and/or radio
evolution for approximately two dozen objects: 2 Type Ib supernovae (SNe), 5
Type Ic SNe, and the rest Type II SNe.  A much larger
list of more than 100 additional SNe have low radio upper limits
(Table 1).

\begin{table}
\begin{center} 
\scriptsize
\begin{tabular}{lcclcclcclcc} 
SN & Type & Radio & SN & Type &
Radio & SN & Type & Radio & SN & Type & Radio \\
\hline
1895B  & I     &    & 1901B  & I     &    & 1909A  & II    &    & 1914A  & ?     &    \\ 
1917A  & I?    &    & 1921B  & II    &    & 1921C  & I     &    & 1923A  & IIP   & DT \\ 
1937C  & Ia    &    & 1937F  & II    &    & 1939C  & II    &    & 1940A  & IIL   &    \\ 
1945B  & ?     &    & 1948B  & II?   &    & 1950B  & II?   & DT & 1954A  & I     &    \\ 
1954J  & II    &    & 1957D  & II?   & DT & 1959D  & II    &    & 1959E  & I     &    \\ 
1963J  & I     &    & 1966B  & II    &    & 1968D  & II    & DT & 1968L  & IIP   &    \\ 
1969L  & IIP   &    & 1970A  & II?   &    & 1970G  & IIL   & LC & 1970L  & I?    &    \\ 
1970O  & ?     &    & 1971G  & I?    &    & 1971I  & Ia    &    & 1971L  & Ia    &    \\ 
1972E  & Ia    &    & 1973R  & IIP   &    & 1974E  & ?     &    & 1974G  & Ia    &    \\ 
1975N  & Ia    &    & 1977B  & ?     &    & 1978B  & II    &    & 1978G  & II    &    \\ 
1978K  & II    & LC & 1979B  & Ia    &    & 1979C  & IIL   & LC & 1980D  & IIP   &    \\ 
1980I  & Ia    &    & 1980K  & IIL   & LC & 1980L  & ?     &    & 1980N  & Ia    &    \\ 
1980O  & II    &    & 1981A  & II    &    & 1981B  & Ia    &    & 1981K  & II?   & LC \\ 
1982E  & Ia?   &    & 1982R  & Ib?   &    & 1983G  & I     &    & 1983K  & IIP/n &    \\ 
1983N  & Ib    & LC & 1984A  & I     &    & 1984E  & IIL/n &    & 1984L  & Ib    & LC \\ 
1984R  & ?     &    & 1985A  & Ia    &    & 1985B  & Ia    &    & 1985F  & Ib    &    \\ 
1985G  & IIP   &    & 1985H  & II    &    & 1985L  & IIL   & DT & 1986A  & Ia    &    \\ 
1986E  & IIL   & LC & 1986G  & Ia    &    & 1986I  & IIP   &    & 1986J  & IIn   & LC \\ 
1986O  & Ia    &    & 1987A  & IIpec & LC & 1987B  & IIn   &    & 1987D  & Ia    &    \\ 
1987F  & IIpec &    & 1987K  & IIb   &    & 1987M  & Ic    &    & 1987N  & Ia    &    \\ 
1988I  & IIn   &    & 1988Z  & IIn   & LC & 1989B  & Ia    &    & 1989C  & IIn   &    \\ 
1989L  & IIL   &    & 1989M  & Ia    &    & 1989R  & IIn   &    & 1990B  & Ic    & LC \\ 
1990K  & IIL   &    & 1990M  & Ia    &    & 1991T  & Ia    &    & 1991ae & IIn   &    \\ 
1991ar & Ic    &    & 1991av & IIn   &    & 1991bg & Ia    &    & 1992A  & Ia    &    \\ 
1992H  & IIP   &    & 1992ad & IIP?  & DT & 1992bd & II    &    & 1993G  & II    &    \\ 
1993J  & IIb   & LC & 1993N  & IIn   &    & 1993X  & II    &    & 1994D  & Ia    &    \\ 
1994I  & Ic    & LC & 1994P  & II    &    & 1994W  & IIn   &    & 1994Y  & IIn   &    \\ 
1994ai & Ic    &    & 1994ak & IIn   &    & 1995G  & IIn   &    & 1995N  & IIn   & DT \\ 
1995X  & IIP   &    & 1995ad & IIP   &    & 1995al & Ia    &    & 1996N  & Ic    & DT \\ 
1996W  & IIpec &    & 1996X  & Ia    &    & 1996ae & IIn   &    & 1996an & II    &    \\ 
1996aq & Ic    &    & 1996bu & IIn   &    & 1996cb & IIb?  & DT & 1997X  & Ic    & DT \\ 
1998bw & Ic    & LC \\
\end{tabular}
\end{center}
\caption{Observed Supernovae (DT = Detection; LC = Light Curve Available)}
\end{table}

In this extensive study of the radio emission from SNe, several
effects have been noted: 1) Type Ia SNe are not radio emitters to the
detection limit of the VLA\footnote{The VLA is operated by the NRAO of
the AUI under a cooperative agreement with the NSF.}; 2) Type Ib/c SNe
are radio luminous with steep spectral indices (generally $\alpha <
-1$; $S \propto \nu^{+\alpha}$) and a fast turn-on/turn-off, usually
peaking at 6 cm near or before optical maximum; and 3) Type II SNe
show a range of radio luminosities with flat spectral indices
(generally $\alpha > -1$) and a relatively slow turn-on/turn-off,
usually peaking at 6 cm significantly after optical maximum. Type Ib/c
may be fairly homogeneous in their radio properties, while Type II, as
in the optical, are quite diverse.

There are a large number of physical properties of SNe which we can
determine from radio observations.  VLBI imaging shows the symmetry of
the explosion and the local CSM, estimates the speed and deceleration
of the SN shock propagating outward from the explosion and, with
assumptions of symmetry and optical line/radio sphere velocities,
allows independent distance estimates to be made (see, \eg
\cite{Marcaide97}, \cite{Bartel85}).

Measurements of the multi-frequency radio light curves and their
evolution with time show the density and structure of the CSM,
evidence for possible binary companions, clumpiness or filamentation
in the presupernova wind, mass-loss rates and changes therein for the
presupernova stellar system and, through stellar evolution models,
estimates of the ZAMS presupernova stellar mass and the stages through
which the star passed on its way to explosion.  It has also been
proposed by Weiler \etal\/(1998) that the time from explosion to 6 cm
radio maximum may be an indicator of the radio luminosity and thus an
independent distance indicator.
 
A summary of the radio information on SNe can be found at 
{\it http://rsd-www.nrl.navy.\\mil/7214/weiler/sne-home.html}.

\section{Models}

All known RSNe appear to share common properties of: 1) nonthermal
synchrotron emission with high brightness temperature; 2) a decrease
in absorption with time, resulting in a smooth, rapid turn-on first at
shorter wavelengths and later at longer wavelengths; 3) a power-law
decline of the flux density with time at each wavelength after maximum
flux density (optical depth $\approx 1$) is reached at that
wavelength; and 4) a final, asymptotic approach of spectral index
$\alpha$ to an optically thin, nonthermal, constant negative value
(\cite{Weiler86}, \cite{Weiler90}). Chevalier (1982a,b) has proposed 
that the relativistic electrons and
enhanced magnetic field necessary for synchrotron emission arise from
the SN shock interacting with a relatively high density circumstellar
medium (CSM) which has been ionized and heated by the initial UV/X-ray
flash.  This CSM is presumed to have been established by a constant
mass-loss ($\dot M$) rate, constant velocity ($w$) wind (\ie $\rho
\propto r^{-2}$) from a red supergiant (RSG) progenitor or companion.
This ionized CSM is also the source of the initial absorption.  A
rapid rise in the observed radio flux density results from the shock
overtaking more and more of the wind material, leaving progressively
less of it along the line of sight to the observer to absorb the more
slowly decreasing synchrotron emission from the shock region.

\subsection{Parameterized Radio Light Curves}

The parameterized model of Weiler \etal\/(1986), Weiler, Panagia, \& Sramek (1990), 
and Montes, Weiler, \& Panagia (1997) may be written as:

\begin{equation}
S({\rm mJy}) = K_1 ~ \left( \frac {\nu} {{\rm 5~GHz}} \right)
^{\alpha} ~ \left( \frac {{t-t_0}} {{\rm 1~day}} \right) ^{\beta}
~ e^{- \left( \tau + \tau^{\prime \prime} \right) } ~ \left( \frac
{1-e^{-\tau^{\prime}}} {\tau^{\prime}} \right) ,
\end{equation}

where

\begin{equation} 
\tau = K_2 ~ \left( \frac {\nu} {{\rm 5~GHz}} \right) ^{-2.1} ~ \left(
\frac {t-t_0} {{\rm 1~day}} \right) ^{\delta},
\end{equation}

\begin{equation}
\tau^{\prime} = K_3 ~ \left( \frac {\nu} {{\rm 5~GHz}} \right) ^{-2.1}
~ \left( \frac {t-t_0} {{\rm 1~day}} \right) ^{\delta ^{\prime}},
\end{equation}

and

\begin{equation}
\tau^{\prime \prime} = K_4 ~ \left( \frac {\nu} {{\rm 5~GHz}} \right)
^{-2.1} .
\end{equation}
\\
\noindent $K_1$, $K_2$, and $K_3$ correspond, formally, to the
unabsorbed flux density ($K_1$), and the uniform ($K_2$) and
non-uniform ($K_3$) optical depths in the surrounding CSM at 5 GHz one
day after the explosion date $t_0$.  $K_4$ represents a non-time
dependent HII absorption along the line-of-sight to the radio emitting 
region (EM = $8.93~\times~10^{7}~K_4~[T_e/10^4 K]^{1.35}$~pc cm$^{-6}$,
where $T_e$ is the electron temperature of the ionized absorbing
region; Eq.~1 -- 223 of Lang 1986). The term $e^{-{\tau}}$ describes the
attenuation of a local medium with optical depth $\tau$ and time
dependence $\delta$ that uniformly covers the emitting source
(``uniform external absorption''); the term
$({1-e^{-\tau^{\prime}}})~{\tau^\prime}^{-1}$ describes the
attenuation produced by an inhomogeneous medium with optical depths
distributed between 0 and $\tau^\prime$ (``clumpy absorption'') and
time dependence $\delta^\prime$; and the term $e^{-{\tau^{\prime\prime}}}$
describes an absorption along the line-of-sight which is sufficiently
far removed from the radio generating region to be constant with time.
All absorbing media are assumed to be purely thermal, ionized hydrogen
with opacity $\propto \nu^{-2.1}$.

A cartoon of the expected structure of the SN and its surrounding
media is presented in Fig.~1.  The radio emission is expected to
arise near the outgoing shock (\cite{Chevalier94}).

\begin{figure}
\vspace{3cm}
\plotfiddle{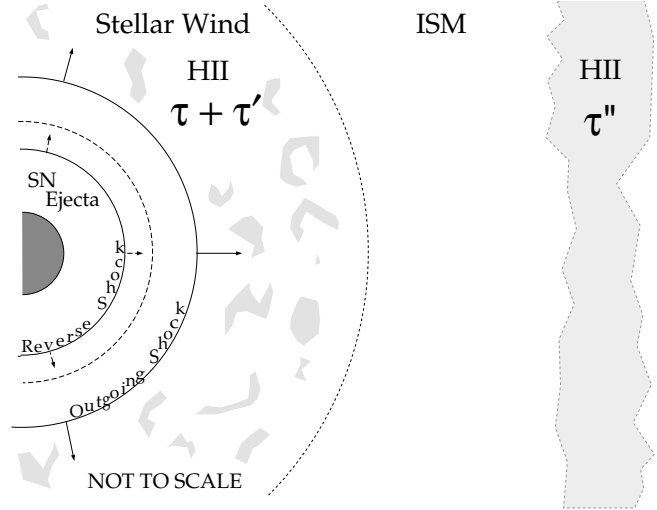}{4.0cm}{-90}{35}{35}{-140}{200}
\caption{Cartoon, not to scale, of the SN and its shocks along with
the stellar wind established CSM and more distant ionized material.
The radio emission is thought to arise near the outgoing shock with
the expected location of the several absorbing terms in Eqs.~2.2 --
2.4 illustrated.}
\end{figure} 

\section{Results}

The success of the basic parameterization and model description can be
seen in the relatively good correspondence between the model fits and
the data for all three subtypes of RSNe, $\eg$ Type Ib SN 1983N
(Fig.~2), Type Ic SN 1990B (Fig.~3), and Type II SN 1979C (Fig.~4a) and
SN 1980K (Fig.~4b).  (Note that after day $\sim4000$, the evolution of the
radio emission from both SN 1979C and SN 1980K deviates from the
expected model evolution; see $\S$4 for discussion of these changes.)

\begin{figure}
\vspace{3cm}
\plotfiddle{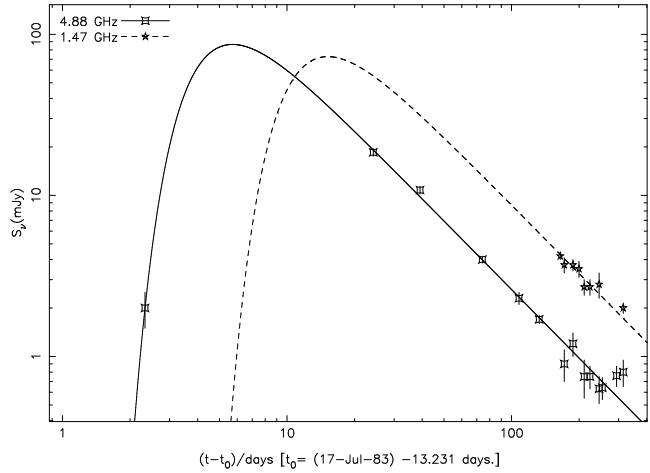}{4.0cm}{-90}{35}{35}{-140}{220}
\caption{Type Ib SN 1983N at 6 cm (4.9 GHz; {\it squares}, 
{\it solid line}) and 20 cm (1.5 GHz; {\it stars}, {\it dashed line}).}
\end{figure} 

\begin{figure}
\vspace{2.75cm}
\plotfiddle{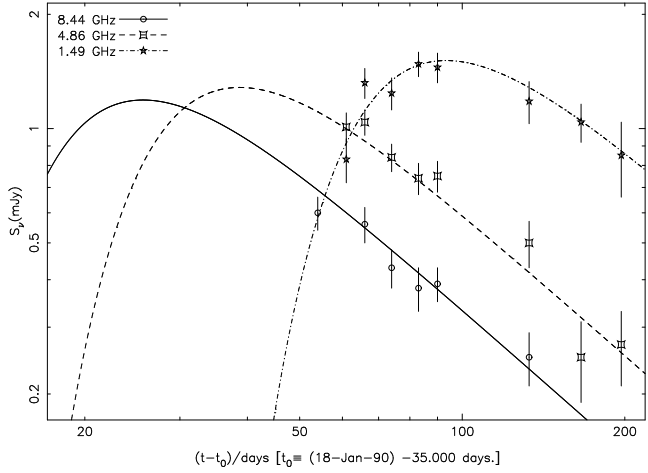}{4.0cm}{-90}{35}{35}{-140}{220}
\caption{Type Ic SN 1990B at 3.4 cm (8.4 GHz; {\it circles}, {\it solid line}), 
6 cm (4.9 GHz; {\it squares}, {\it dashed line}), and 20 cm (1.5 GHz; {\it stars}, {\it dash-dot line})}.
\end{figure} 

\begin{figure}
\vspace{1cm}
\plotfiddle{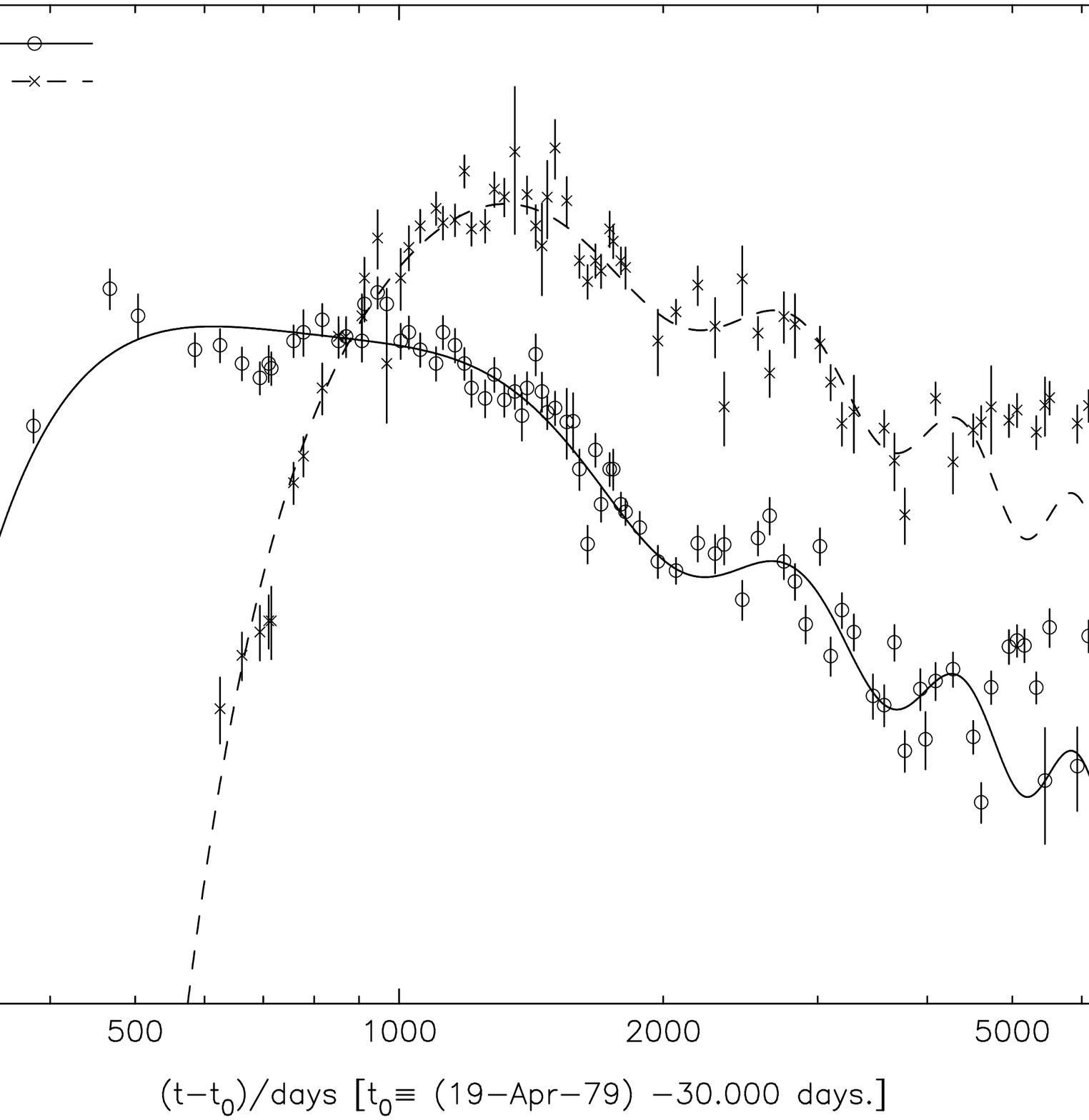}{4.0cm}{0}{22}{22}{-160}{0}
\caption{Shown are the radio light curves of Type II SN 1979C (Fig.~4a, left)
at 6 cm ({\it circles}) and 20 cm ({\it crosses}) as observed over a period of
about 18 years together with the best-fit model curves ({\it solid} and
{\it dashed} curves, respectively).  The model light curves for SN 1979C
include the effects due to a possible binary companion in an 
eccentric orbit (Weiler \etal\/1991, 1992a). Note the {\it increase} in flux 
density of
SN 1979C, with respect to the expected, continuing decline, starting
at $\sim$4300 days (Montes \etal\/2000).  Also shown, with the same
notation at the same frequencies, are the radio light curves of
Type II SN 1980K (Fig.~4b, right).  Note the sudden {\it drop} in the flux
density of SN 1980K, with respect to the expected, continuing decline,
which occurs at $\sim$3700 days (Montes \etal\/1998). }
\plotfiddle{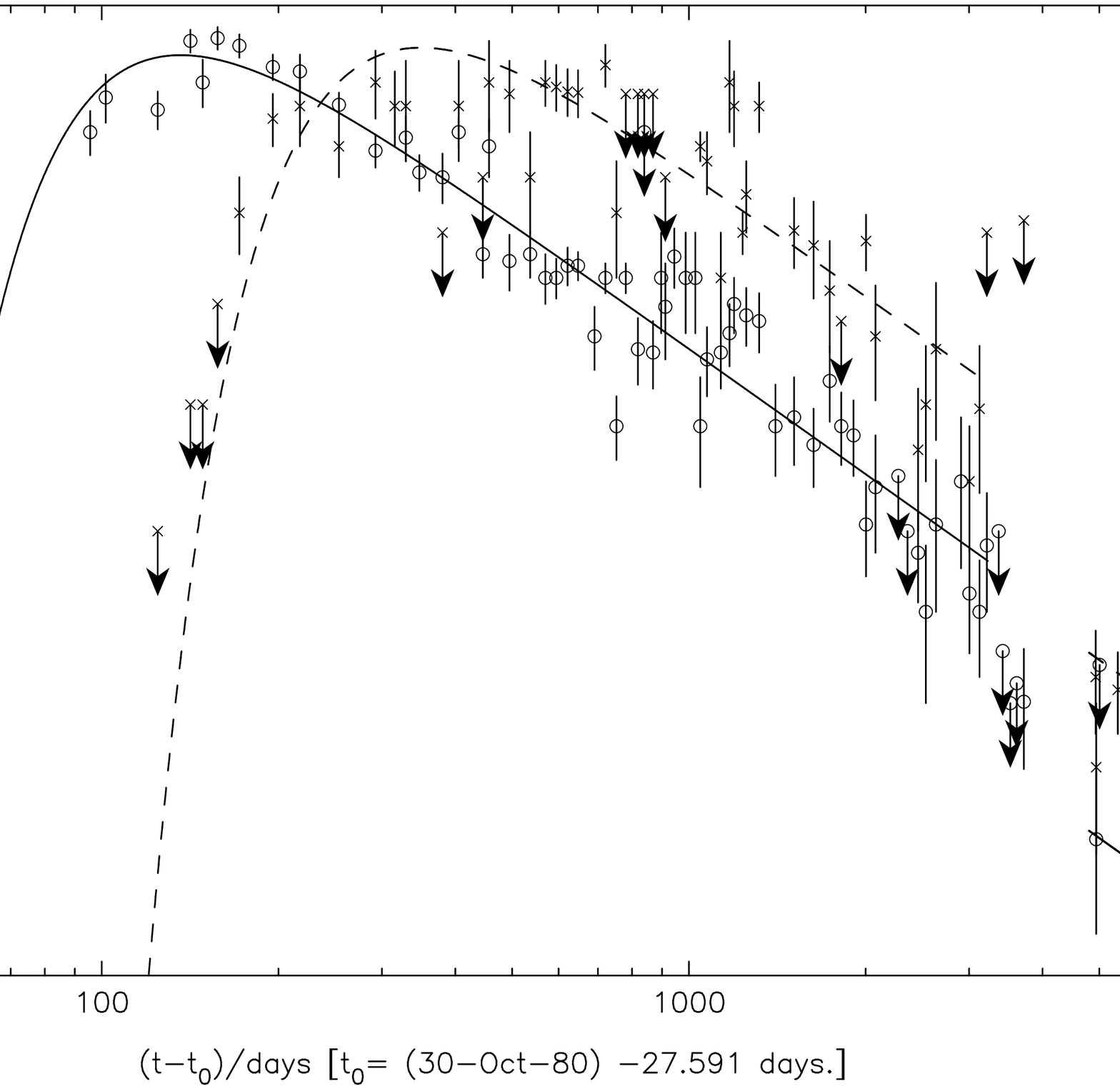}{0cm}{0}{23}{23}{40}{118}
\end{figure}

\subsection{Mass-loss Rate \& Change in Mass-loss Rate}

From the Chevalier (1982a,b) model, the turn-on of the radio emission
for RSNe provides a measure of the presupernova mass-loss-rate to 
stellar-wind-velocity ratio.  Using the formulation of Weiler \etal\/(1986; Eq.~16),
we can write

\begin{equation}
\frac{\dot M (M_\odot ~ {\rm yr}^{-1})}{( w / 10\ {\rm km\ s}^{-1} )} =
3 \times 10^{-6}\ K_2^{0.5}\ m^{-1.5} {\left(\frac{v_i}{10^{4}\ {\rm km\
s}^{-1}}\right)}^{1.5} {\left(\frac{1}{45\ {\rm days}}\right) }^{1.5
m} {\left( \frac{T}{10^{4}\ {\rm K}} \right)}^{0.68}
\end{equation}

\noindent where $\dot M$ is the presupernova mass-loss rate, $w$ is
the presupernova wind velocity, $K_2$ is the same as in Eq.~2.2, $m$ is
the SN shock deceleration index (shock  radius $\propto~t^m$),
$v_i$ is the initial SN shock velocity ($\sim13,000$ \kms),
and $T$ is the temperature of the circumstellar material ($\sim30,000$ K).

From Eq.~3.5 the mass-loss rates from SN progenitors are generally
estimated to be $\sim10^{-6}$ M$_\odot$ yr$^{-1}$ for Type Ib/c SNe
and $\sim10^{-4}$ M$_\odot$ yr$^{-1}$ for Type II SNe.  For the
specific case of SN 1993J, where detailed radio observations are
available starting just a few days after explosion, Van Dyk \etal\/(1994) 
find evidence for a changing mass-loss rate (Fig.~5) for the
presupernova star, which was as high as $\sim10^{-4}$ M$_\odot$
yr$^{-1}$ approximately 1000 years before explosion and decreased to
$\sim10^{-5}$ M$_\odot$ yr$^{-1}$ just before explosion. Recently
Fransson \& Bj\"orgsson (1998) have shown that the observed behavior of the free-free
absorption for SN 1993J could alternatively be explained in terms of a
systematic decrease of the electron temperature in the circumstellar
material as the SN expands.  It is not clear, however, what the
physical process is which determines why such a cooling occurs
efficiently in some SNe and not in others.

\begin{figure}
\vspace{3cm}
\plotfiddle{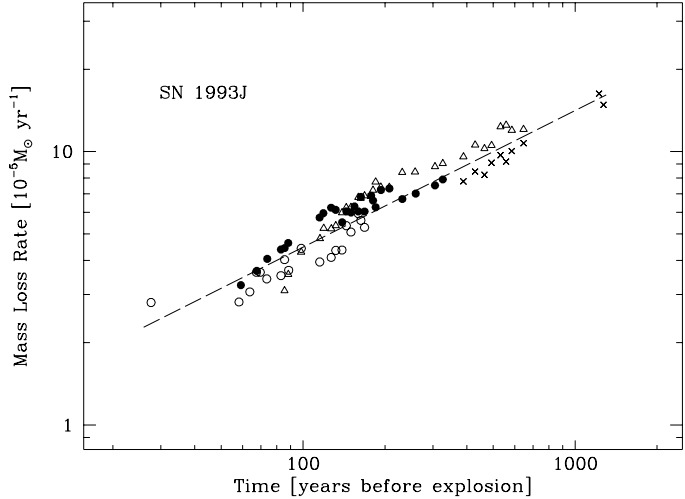}{4.0cm}{0}{35}{35}{-45}{-45}
\caption{Mass-loss rate of the presumed red supergiant progenitor to
SN 1993J {\it vs.} time before the explosion (Van Dyk \etal\/1994).}
\end{figure}

\subsection{Clumpiness of the Presupernova Wind}
 
In their study of the radio emission from SN 1986J, Weiler, Panagia, \& Sramek 
(1990) found that the simple Chevalier (1982a,b) model
could not describe the relatively slow turn-on.  They therefore added
terms described mathematically by $\tau^\prime$ in Eqs.~2.1 and 2.3.  This
extension greatly improved the quality of the fit and was interpreted
by Weiler, Panagia, \& Sramek (1990) to represent the possible presence of
filamentation or clumpiness in the CSM.

Such a clumpiness in the wind material was again required for modeling
the radio data from SN 1988Z (\cite{Vandyk93}) and SN 1993J
(\cite{Vandyk94}).  Since that time, evidence for filamentation in the
envelopes of SNe has also been found from optical and UV observations
(\cite{Filippenko94}, \cite{Spyromilio94}).

\subsection{Binary Systems}

In the process of analyzing a full decade of radio measurements from
SN 1979C, Weiler \etal\/(1991, 1992a) found evidence for a
significant, quasi-periodic, variation in the amplitude of the radio
emission at all wavelengths of $\sim15$~\% with a period of 1575 days
or $\sim4.3$ years (see Fig.~4a at age $<$4300 days).  They
interpreted the variation as due to a minor ($\sim8$~\%) density
modulation, with a period of $\sim4000$ years, on the larger,
relatively constant presupernova stellar mass-loss rate.  Since such a
long period is inconsistent with most models for stellar pulsations,
they concluded that the modulation may be produced by interaction of a
binary companion in an eccentric orbit with the stellar wind from the
presupernova RSG.  This concept was strengthened by more detailed 
calculations for a binary model from Schwarz \& Pringle (1996).
 Since that time, the presence of binary companions has been suggested
for the progenitors of SN 1993J (\cite{Podsiadlowski93}) and SN 1994I
(\cite{Nomoto94}), indicating that binaries may be common in
presupernova systems.

\subsection{HII Along the Line-of-Sight}

A reanalysis of the radio data for SN 1978K from Ryder \etal\/(1993)
clearly shows flux density evolution characteristic of normal Type II
SNe.  Additionally, the data indicate the need for a time-independent,
free-free absorption component along the line-of-sight.
Montes, Weiler, \& Panagia (1997) interpret this constant absorption term as indicative
of the presence of HII along the line-of-sight to SN 1978K,
perhaps a part of an HII region or a distant circumstellar shell
associated with the SN progenitor. SN 1978K had already
been noted for its lack of optical emission lines broader than a few thousand
\kms\/ since its discovery in 1990 (\cite{Ryder93}), indeed suggesting
the presence of slowly moving circumstellar material.

To determine the nature of this absorbing region, a high-dispersion
spectrum of SN 1978K at the wavelength range 6530 -- 6610 \AA~ was
obtained by Chu \etal\/(1999). The spectrum shows not only the
moderately broad H$\alpha$ emission of the supernova ejecta, but also
narrow nebular H$\alpha$ and [N II] emission. The high [N II]
6583/H$\alpha$ ratio of 0.8 -- 1.3 suggests that this radio absorbing
region is a stellar ejecta nebula. The expansion velocity and emission
measure of the nebula are consistent with those seen in ejecta nebulae
of luminous blue variables. Previous low-dispersion spectra have
detected a strong [N II] 5755\AA~ line, indicating an electron density
of (3 -- 12) $\times 10^5$~cm$^{-3}$. These data suggest that the ejecta
nebula detected towards SN 1978K is probably part of the pre-shock
dense circumstellar envelope of SN 1978K. Another possible example of
this type of system may be SN 1997ab, which looks in its optical
spectrum like a young version of SN 1978K.

\section{Rapid Pre-Supernova Stellar Evolution} 

SN radio emission that preserves its spectral index while deviating
from the standard model is taken to be evidence for a change of the
average circumstellar density behavior from the canonical $r^{-2}$ law expected
for a pre-SN wind with a constant mass-loss rate, $\dot M$, and a
constant wind velocity, $w$. Since the radio luminosity of a SN is
proportional to $({\dot M}/w)^{(\gamma-7+12m)/4}$
(\cite{Chevalier82a}) or, equivalently, to the same power of the
circumstellar density (since $\rho_{\rm CSM} \propto {\dot M}/w$), a
measure of the deviation from the standard model provides an
indication of deviation of the circumstellar density from the $r^{-2}$
law.  Monitoring the radio light curves of RSNe also provides a rough
estimate of the time scale of deviations in the presupernova stellar
wind density.  Since the SN shock travels through the CSM roughly 1000
times faster than the stellar wind velocity which established the CSM
($v_{\rm shock} \sim10,000$ \kms~{\it vs.}~$w_{\rm wind} \sim10$ \kms) one
year of radio light curve monitoring samples roughly 1000 years of
stellar wind mass-loss history.

\subsection{Radio Evidence for CSM Structure}

{\bf SN 1979C} (Type IIL) prior to 1991 (age $<$4300 days; $\sim$12
years) follows a standard, albeit sinusoidally modulated, declining
radio emission (see $\S$3.3). However, for age $>$4300 days a slow
increase in the radio light curve occurs at all wavelengths (see
Fig.~4a). By day $\sim7100$, this change in evolution implies an {\it
excess} in flux density by a factor of $\sim$1.7 with respect to the
standard model, or a density enhancement by a factor of $\sim$1.34
over the expected density at that radius. This may be understood as a
change of the average CSM density profile from the $r^{-2}$ law which
was applicable until day $\sim$4300, to an appreciably flatter
behavior of $\sim r^{-1.4}$ (\cite{Montes00}).
 
{\bf SN 1980K} (Type IIL) prior to epoch $\sim$3700 days ($\sim10$
years) is also well behaved.  However, more recent measurements show a
steep {\it decline} in flux density at all wavelengths by a factor of
$\sim2$ occurring between day $\sim$3700 and day $\sim$4900 (see
Fig.~4b).  Such a sharp decline in flux density implies a {\it decrease} in
$\rho_{\rm CSM}$ by a factor of $\sim1.6$ below that
expected for a $r^{-2}$ CSM density profile (\cite{Montes98}).

{\bf SN 1988Z} (Type IIn), similarly to SN 1980K, shows a sharp drop
in its flux density with respect to its expected radio evolution at an
age of a few thousand days (several years).  Although the parameters
of the change are yet to be quantified, it appears to also have
evolved rapidly in the last several thousand years before explosion
(\cite{Lacey00}).

{\bf SN 1987A} (Type II) is the best studied RSN because its proximity
makes it easily detectable even at very low radio brightness.  The
progenitor to SN 1987A was in a blue supergiant (BSG) phase at the
time of explosion and had ended a RSG phase some ten thousand years
earlier.  After an initial, very rapidly evolving radio outburst
(\cite{Turtle87}) which reached a peak flux density at 6 cm $\sim3$
orders-of-magnitude fainter than other known Type II RSNe (presumably
due to sensitivity limited selection effects), the radio emission
declined to a low radio brightness within a year.  However, at an age
of $\sim3$ years the radio emission started increasing again and
continues to increase at the present time (see \cite{Ball95},
\cite{Gaensler97}).  Although its extremely rapid development
resulted in the early radio data at higher frequencies being almost
non-existent, the evolution of the initial radio outburst is roughly
consistent with the models described above in Eqs.~2.1 -- 2.4 (\ie a
shock front expanding into a spherically symmetric circumstellar
envelope).  The density implied by such modeling is appropriate to a
pre-SN mass-loss rate of $\sim10^{-7}$ M$_\odot$ yr$^{-1}$ for a
wind velocity of $\sim150$ \kms.  Because the {\sl Hubble Space Telescope
(HST)\/} can actually
image the denser regions of the CSM around SN 1987A, we know that the
current rise in radio flux density is caused by the interaction of
the SN shock with the diffuse material at the inner edge of the well
known inner circumstellar ring (\cite{Gaensler97}). Since the density
increases as the SN shock interaction region moves deeper into the
main body of the optical ring, the flux density is expected to
continue to increase steadily at all wavelengths.  Recently, increases
at optical and X-ray have also been reported (\cite{Garnavich97},
\cite{Hasinger96}).  Best estimates are that the shock/CSM
interaction will reach a maximum by $\sim2003$.

\subsection{Discussion of CSM Structure} 

For at least four supernovae, namely SN 1979C, SN 1980K, SN 1988Z, and
SN 1987A, we have significant changes in radio flux
density occurring a few years after the explosion. Since the SN shock
is moving about 1000 times faster than the wind material of the RSG
progenitor (\ie $\sim10,000$ \kms ~{\it vs.} $\sim10$ \kms), such a
time interval implies a significant change in the pre-supernova
stellar wind properties several thousand years before the explosion.
Such an interval is short compared to the lifetimes of typical RSN
progenitors (say, 10 -- 30 Myrs) but is a sizeable fraction of its red
supergiant phase ($t_{\rm RSG} \sim$2 -- 5 $\times 10^5$ yrs),
suggesting that a significant transition occurs in the evolution of
pre-supernova stars just before the final explosive event.

Since the radio emission is determined by the mass-loss-rate to 
stellar-wind-velocity ratio (${\dot M}/w$), one of these quantities, or
both are required to change by as much as a factor of 2 over the last
few thousand years before the SN explosion.  Such a time is too short
(for H and He burning), or too long (for C and heavier element
burning) to correspond to any of the known nuclear burning phases and,
therefore, it is unlikely that the stellar luminosity (which
determines the mass-loss rate, ${\dot M} \propto L^{1-1.5}$) can
vary on a time scale needed to account for the observed changes.

On the other hand, the wind velocity, $w$, is roughly proportional to
the square of the effective temperature ($w \propto T_{\rm eff}^2$, \eg
\cite{Panagia82}) so that a change of a factor of $\sim2$ in
$w$ requires a change of a factor of only $\sim1.4$ in $T_{\rm eff}$, \eg
from $\sim3,500$ K to $\sim5,000$ K or, correspondingly, a change from
an early M to an early K supergiant spectrum.  Such a transition would
define a loop in the HR diagram reminiscent of the blue loops which
are characteristic of the evolution of moderately massive stars (\eg
\cite{Brocato93}, \cite{Langer95}).  However, the
apparent transition implied by these CSM density changes cannot be
classical blue loops, since classical blue loops are much slower and more extreme
processes occurring several $\times 10^5$ years before the terminal
stages of an RSG and involving temperature excursions from $\sim3,500$
K to $>10,000$ K.

The smaller temperature changes which we infer from the radio data
require a star to change only from a very red to a moderately red
spectrum, and back, corresponding to a transition in the HR diagram
which is more appropriately dubbed a {\it ``pink loop."} The cause of
such loops is not obvious, but may be similar to the not-so-well
understood phenomenon that caused the SN 1987A progenitor to move in
the HR diagram from being a red supergiant to a blue supergiant some
$10^4$ years before explosion.

Another possibility for explaining these implied CSM density changes
around at least some presupernova stars derives from a recent study by
Panagia \& Bono (this Conference).  They find from modeling that the pulsational
instability of stars in the mass range 10 -- 20 ${M_\odot}$ may, in some
cases, be of suitable period and magnitude to account for the changes
of the pre-supernova mass-loss rates implied by radio observations of
RSNe.

\begin{figure}
\vspace{3cm}
\plotfiddle{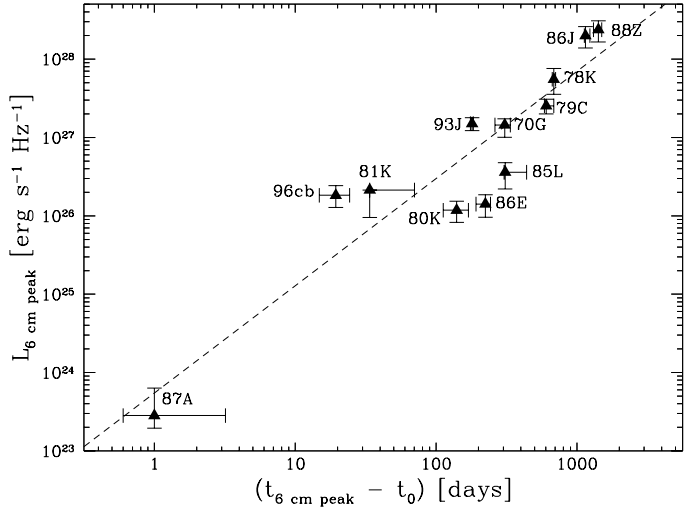}{4.0cm}{-90}{35}{35}{-135}{215}
\caption{Peak 6 cm luminosity, $L_{\rm 6\ cm\ peak}$, of RSNe {\it
vs.\/} time, in days, from explosion to peak 6 cm flux density
($t_{\rm 6\ cm\ peak} - t_0$) for Type II SNe.  The {\it dashed
line} (given by Eq.~5.6 in the text) is the unweighted, best fit to
the 12 available Type II RSNe.  Error bars are based on best
estimates.  Where no error or only a stub of a line is shown, the
error in that direction is indeterminate (from Weiler \etal\/1998).}
\end{figure} 

\section{Peak Radio Luminosities and Distances}

Our long-term monitoring of the radio emission from supernovae shows
that the radio ``light curves'' evolve in a systematic fashion with a
distinct peak flux density (and thus, in combination with a distance,
a peak spectral luminosity) at each frequency and a well-defined time
from explosion to that peak.  Studying these two quantities at 6 cm
wavelength, peak spectral luminosity ($L_{\rm 6\ cm\ peak}$) and time
after explosion date ($t_0$) to reach that peak ($t_{\rm 6\ cm\
peak}-t_0$), we find that they appear related (Fig.~6; see also
\cite{Weiler98}).  In particular, based on twelve objects, Type II
supernovae appear to obey a relation

\begin{equation}
L_{\rm 6\ cm\ peak} \simeq 5.5 \times 10^{23}~ (t_{\rm 6\
cm\ peak} - t_0)^{1.4}~ {\rm erg~s^{-1} Hz^{-1}}
\end{equation}

\noindent
with time measured in days. Thus, if this relation is supported by
further observations, it provides a means for determining distances to
supernovae, and thus to their parent galaxies, from purely radio
continuum observations.

Although there are still relatively few objects to which this
techniques can be applied, RSNe could eventually provide a powerful
and independent technique for investigating the long-standing problem
of distance estimates in astronomy.  With such intrinsically bright
Type II RSNe as SN 1988Z and SN 1986J, the technique can be applied to
distances of at least 100 Mpc with current VLA technology.  With
future sensitivity improvements and planned, new, more sensitive radio
telescopes, the technique could be extended to large distances, even
for less luminous RSNe.  For example, with a sensitivity of 1 $\mu$Jy,
a SN of the same class as SNe~1988Z and 1986J could be detected to a
redshift of $\sim1$, while a relatively radio faint Type II SN, such
as SN 1980K, could be studied to a redshift of $>0.1$.

\begin{figure}
\vspace{3cm}
%\plotfiddle{kweiler.fig7.ps}{13.25cm}{0}{75}{75}{-225}{-60}
\caption{VLBI radio images of SN 1993J at 6 cm wavelength.  Images on
the left-hand side are normalized to the same peak brightness to
emphasize structural changes.  Images on the right-hand side are on a
single brightness scale to illustrate the decrease in brightness with
time (from Marcaide \etal\/1997).}
\end{figure}

\section{Sphericity of an SN explosion}

It has often been suggested that SN explosions are non-spherical, and
there is evidence in a number of stellar systems for jets, lobes, and
other directed mass-loss phenomena.  Also, the presence of
polarization in the optical light from SNe (including SN 1993J) has
been interpreted for non-sphericity (see, \eg \cite{Hoeflich96})
and probably the most obvious evidence for non-spherical structure in
an SN system is the very prominent inner ring around SN 1987A.
However, our most direct evidence for the structure of at least the
shock wave from an SN explosion and the CSM with which it is
interacting is from VLBI measurements on SN 1993J.  A series of images
taken by Marcaide and co-workers (Fig.~7; \cite{Marcaide97}) over a
period of two years from 1994 September through 1996 October show only
a very regular ring shape indicative of a relatively spherical
shock wave expanding into a relatively uniform CSM.  The cause of such
apparently conflicting results is still to be resolved.

\section{Summary of RSN Studies}

Arising from one of the most energetic phenomena in the Universe, the
radio emission from supernovae appears to be relatively well
understood in terms of shock interaction with a structured
circumstellar medium as described by the Chevalier (1982a,b) model and
its modifications (\cite{Weiler86}, \cite{Weiler90}, and \cite{Weiler91}).  
With this
modeling, the radio emission can be used to estimate the circumstellar density, 
the pre-SN mass
loss rate and changes therein, to show the existence of filamentation
in the pre-SN stellar wind, to indicate the possible presence of
binary companions, and to measure the symmetry of the explosion and
the CSM with which it is interacting.  More speculatively, radio
observations may also lead to a more physical classification system
for SNe and provide a new technique for estimating distances to SNe
and their parent galaxies.

However, the recent suggestion of an association of the $\gamma$-ray
burst GRB~980425 with the Type Ic supernova SN 1998bw provides
evidence for yet a new phenomenon which may arise in at least some
types of SN explosions.  Since SN 1998bw is a strong radio emitter, it
is important to consider how it compares with the known properties of
RSNe discussed above.
 
\section{Gamma-Ray Bursts and SN 1998bw}

\subsection{Gamma-ray Burst Afterglows}

Gamma-ray bursts (GRBs) are ``mysterious" flashes of high-energy
radiation that appear from random directions in space and typically
last a few seconds. They were first discovered by U.S.~Air Force Vela
satellites in the 1960s and, since then, numerous theories of their
origin have been proposed.  NASA's Compton Gamma-Ray
Observatory (CGRO) satellite has detected several thousand bursts so
far, with an occurrence rate of approximately one per day.  The
uniform distribution of the bursts on the sky has led theoreticians to
suggest that their sources are either very near, and thus uniformly
distributed around the solar system, are in an unexpectedly large halo
around the Galaxy, or are at cosmological distances -- not very
restrictive proposals.

The principal limitation to understanding the origin of the bursts has
been the difficulty in pinpointing their direction on the sky in order
to obtain the multi-wavelength observations necessary to constrain
physical models.  Gamma-rays are exceedingly difficult to focus on to a
position sensitive detector, and the bursts' short duration
exacerbates the problem. Only with the launch of the Italian/Dutch
satellite BeppoSAX in 1996 has it been possible to couple a quick
response pointing system with relatively high precision position
sensitive detectors for $\gamma$-rays and hard X-rays.  This quick
response, high accuracy position information has finally permitted
rapid and accurate follow-up observations with the world's powerful
ground-based and space-based telescopes, and has led to the discovery 
of long-lived
``afterglows" of the bursts in soft X-rays, visible and infrared light,
and radio waves. Although the $\gamma$-ray bursts generally last only
seconds, their afterglows have, in a few cases, been studied for
minutes, hours, days, or even weeks after discovery.  These longer
wavelength observations have allowed observers to probe the immediate
environment of $\gamma$-ray burst sources and to assemble clues as to
their nature.

The first GRB related optical transient was identified for GRB~970228
by Groot \etal\/(1997) and followup with \hst\/ by Sahu \etal\/(1997) 
demonstrated that the GRB was associated
with a faint (thus probably distant) late-type galaxy. A few months later
Fruchter \& Bergeron 1997; see also \cite{Pian98}) imaged the
afterglow of another $\gamma$-ray burst, GRB~970508, with the \hst\/ {\it
WFPC2\/} finding this source to be associated with a late-type
galaxy at a redshift of $z = 0.835$.  GRB~970508 was also the first GRB
to be detected in its radio afterglow (\cite{Frail97}).

Heise \etal\/(1997) detected the very energetic $\gamma$-ray burst GRB
971214 on 1997 December 14 with the BeppoSAX satellite and provided
sufficient positional accuracy for Halpern \etal\/(1997) to identify
a visible light afterglow using the KPNO 2.1 m telescope.  As the
visible light from the burst afterglow faded, an extremely faint
galaxy ($R=25.6\pm0.2$) was detected at its position.  Using the
 10-meter Keck II telescope on Mauna Kea, Hawaii,
Kulkarni \etal\/(1998a) measured a redshift of $z = 3.42$.  Subsequent
images taken with the \hst~(\cite{Odewahn98})
confirmed the association of the burst afterglow and the redshift of
this faint galaxy. For such a large distance, Odewahn \etal\/(1998) 
estimated that the amount of energy released in the $\gamma$-ray flash
was extremely high, $\sim3 \times 10^{53}$ erg. Thus, it appears
that at least some GRB sources are very energetic explosions
(hypernovae?; \cite{Paczynski98}, \cite{Iwamoto98}) occurring at
cosmological distances.

\subsection{GRB~980425 and SN 1998bw}

\subsubsection{Background}

While still generally accepted that ``most'' GRBs are extremely
distant and energetic, the discovery of GRB~980425 (\cite{Soffitta98})
on 1998 April 25.90915 and its possible association with a bright
supernova, SN 1998bw at RA(J2000) = $19^h 35^m 03\fs31$,
Dec(J2000) = $-52\arcdeg 50\arcmin 44\farcs7$ (Tinney \etal\/1998), 
in the relatively nearby spiral galaxy ESO~184-G82 at $z =
0.0085$ (distance $\sim38$ Mpc for $H_0 = 65$ \kms\/ Mpc$^{-1}$;
Galama \etal\/1998, \cite{Lidman98}, \cite{Tinney98}, \cite{Sadler98}),
has introduced the possibility of multiple origins for GRBs.  The
estimated explosion date of SN 1998bw in the interval 1998 April 21 --
27 (\cite{Sadler98}) corresponds rather well with the time of
GRB~980425.  Iwamoto \etal\/(1998) feel that they can restrict the core
collapse date for SN 1998bw even more from hydrodynamical modeling of
exploding C + O stars and, assuming that the SN 1998bw optical light
curve is generated by $^{56}$Ni as in Type Ia SNe, they then restrict
the coincidence between the core collapse of SN 1998bw to within
+0.7/$-$2 days of the detection of GRB~980425.

Classified initially as an SN optical Type Ib (\cite{Sadler98}), then
Type Ic (Patat \& Piemonte 1998), then peculiar Type Ic (Kay, 
Halpern, \& Leighly 1998, 
Filippenko 1998), then later, at an age of ~300 - 400 days, again
as a Type Ib (\cite{Patat99}), SN 1998bw presents a number of optical
spectral peculiarities which strengthen the suspicion that it may be
the counterpart of the $\gamma$-ray burst.

However, some doubt remains concerning the association of GRB~980425
with SN 1998bw.  When the more precise BeppoSAX NFI was pointed at the
BeppoSAX error box 10 hours after the detection of GRB~980425, two
X-ray sources were present (\cite{Pian99}).  One of these, named S1
by Pian \etal\/(1999), is coincident with the position of SN 1998bw
and declined slowly between 1998 April and 1998 November.  The
second X-ray source, S2, which was $\sim4\arcmin$ from the position
of SN 1998bw, was, at best, only marginally detected at $<3\sigma$
six days after the initial detection, and not detectable
again in follow up observations in 1998 April, May, and November
(\cite{Pian99}). Even though the {\it a posteriori} statistics
indicate a very low probability($\sim10^{-4}$) of a GRB being nearly
coincident in space and time with a SN outburst, the concern remains
that the Pian source S2 was the brief afterglow from GRB~980425 rather
than the Pian source S1 associated with SN 1998bw.

\subsubsection{Radio emission}

Since the peak absolute magnitude of Type Ib/c SN 1998bw, while bright at $M_B$
$\sim-19.5$ is not much brighter than Type Ib SN 1966J ($M_{\rm B} \sim-19$,
converted from $H_0 = 75$ \kms\/ Mpc$^{-1}$ to $H_0 = 65$ \kms\/ Mpc$^{-1}$, 
\cite{Miller90}) and less luminous than the
Type Ic SN 1992ar, with $M_B \sim -20$ (\cite{Hamuy92}), much of the
argument for the unusual nature of SN 1998bw rests on the radio
observations.  (N.B.:  SN 1966J may have been misclassified.   
\cite{Vandyk96} indicate in a {\it Note Added in Proof} that D. Branch 
has reclassified SN 1966J as a Type Ia.)

The radio emission from SN 1998bw reached an unusually high 6 cm
spectral luminosity at peak of $\sim8 \times 10^{28}$ erg s$^{-1}$
Hz$^{-1}$, \ie about 3 times higher than any of the relatively well
studied radio RSNe and $\sim$ 30 -- 40 times higher than a typical
Type Ib/c SN at peak (see Table 2 and \cite{Weiler98}).  Unfortunately, 
it is not clear how significant is this ``excess'' radio luminosity, 
since only 6
other examples of radio emitting Type Ib/c SNe are known (see Table
1). Also, one must keep in mind that, contrary to the often stated
opinion that SN 1998bw is ``the most luminous radio supernova ever
observed,'' SN 1998bw is still exceeded in peak 6 cm spectral
luminosity by the poorly studied, presumed supernova, SN 1982aa
(\cite{Green94}) in the starburst galaxy NGC 6052 (Markarian 297).
SN 1982aa is estimated to have peaked at a 6 cm spectral luminosity of
$\sim1 \times 10^{29}$ erg s$^{-1}$ Hz$^{-1}$. Although SN 1982aa
was not optically identified and, therefore, has no optical spectral
type classification, its radio evolution strongly resembles that of
Type II RSNe (\cite{Yin94}).

\begin{table}
\begin{center} 
\begin{tabular}{lcc} 
SN Name & Type & Peak 6cm Luminosity \\
        &      & (erg s$^{-1}$ Hz$^{-1}$) \\
\hline
1983N  & Ib    &  $1.4 \times 10^{27}$ \\ 
1984L  & Ib    &  $2.6 \times 10^{27}$ \\ 
1990B  & Ic    &  $5.6 \times 10^{26}$ \\  
1994I  & Ic    &  $1.4 \times 10^{27}$ \\
1997X  & Ic    &  $2.6 \times 10^{27}$ \\
1998bw & Ib/c  &  $7.9 \times 10^{28}$ \\
\end{tabular}
\end{center}
\caption{Peak Radio Luminosities of Type Ib/c Supernovae)}
\end{table}

As a further check for unusual characteristics of SN 1998bw, we can
compare a smoothed 6 cm light curve for SN 1998bw with model 6 cm light
curves for other RSNe.  This is illustrated in Fig.~8.  Inspection
of the figure shows that SN 1998bw is unusual in its radio emission,
but not extreme.  For example, the time from explosion to peak 6 cm
luminosity for both SN 1987A and SN 1983N was faster and, as mentioned
above, the 6 cm spectral luminosity of SN 1998bw at peak is exceeded
by that of SN 1982aa.  However, as is clear from Fig.~8, SN 1998bw is
the most luminous Type Ib/c RSN ever observed by a factor of $\sim30$,
and it reached a higher radio luminosity earlier than any RSN known.

\begin{figure}
\vspace{2.75cm}
\plotfiddle{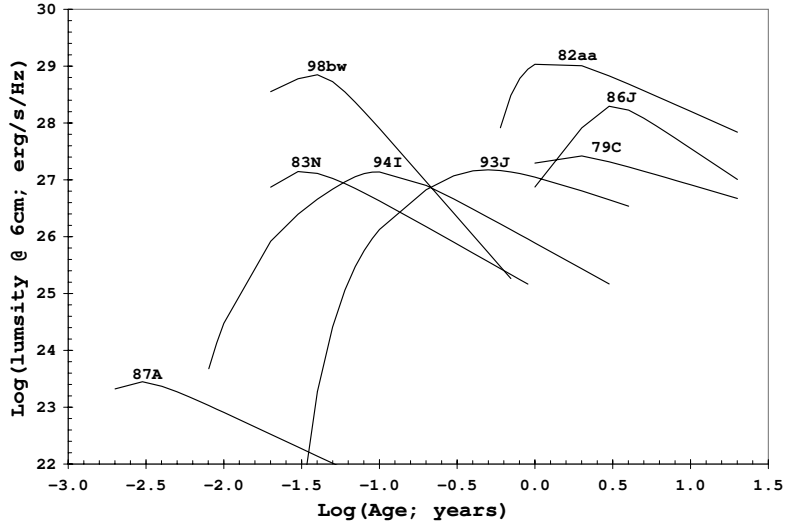}{4.0cm}{-90}{45}{45}{-180}{240}
\caption{Approximate model 6 cm luminosity vs.~time ``light'' curves
for radio supernovae (RSNe).  Derived from best fits to observations,
these model curves smooth out the many ``bumps and wiggles'' seen in
actual data to show the general character of the time evolution rather
than the details.  Both Type Ib/c (SN 1983N, SN 1994I, SN 1998bw) and
Type II (SN 1979C, SN 1982aa, SN 1986J, SN 1987A, SN 1993J) SNe are
shown.}
\end{figure} 

\subsubsection{Brightness temperature}

Although unique in neither the speed of evolution nor in radio
luminosity, SN 1998bw is certainly unusual in the combination of these
two factors -- very radio luminous very soon after explosion.
Kulkarni \etal\/(1998b,c) have used these observed qualities, together with
the lack of interstellar scintillation at early times, brightness
temperature estimates, and physical arguments to conclude that the
shock wave from SN 1998bw giving rise to the radio emission must have
been expanding relativistically.  On the other hand, Waxman \& Loeb (1999) 
argue that a sub-relativistic shock can generate the observed radio
emission.  However, both sets of authors agree that a very high
expansion velocity ($\ge0.3c$) is required for the radio emitting
region under a spherical geometry.

Simple arguments confirm this high velocity since, to avoid the well-known 
``Compton Catastrophe,'' Kellermann \& Pauliny-Toth (1969) have shown
that $T_{\rm B} < 10^{12}$ K must hold, and Readhead (1994) has better
defined this limit to $T_{\rm B} < 10^{11.5}$ K.  From geometrical
arguments, such a limit requires the radiosphere of SN 1998bw
to have expanded at $\ge200,000$ \kms, at least during the
first few days after explosion.  While such a value is still
only mildly relativistic ($\gamma \sim$ 1.5), it seems very high.  
However, measurements by
Gaensler \etal\/(1997) have demonstrated that the radio emitting
regions of SN 1987A have expanded at an {\it average} velocity of
$\sim35,000$ \kms\/ over the 3 years from 1987 February to
mid-1990, so that, in a very low density environment such as one finds
around Type Ib/c SNe, very high shock velocities may be possible.

This is illustrated graphically in Fig.~9 where, for an adopted brightness
temperature of $10^{11.5}$ K, the lines of constant expansion
velocity are shown on a plot of the 6 cm spectral luminosity at peak 
{\it vs.}~the time to reach that peak for a number of RSNe.  Although expected,
it is noteworthy that all Type II SNe, which have relatively dense
circumstellar envelopes, have the lowest expansion velocities, while the
Type Ib/c SNe, with their relatively less dense CSM, have significantly
higher expansion rates.  Once again, SN 1998bw appears to
represent a more extreme form of Type Ib/c RSNe.

\begin{figure}
\vspace{2.75cm}
\plotfiddle{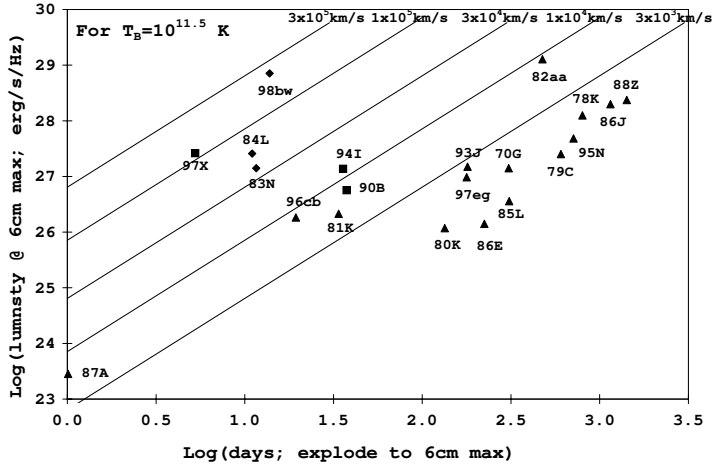}{4.0cm}{90}{40}{40}{160}{-20}
\caption{A plot of the 6 cm luminosity at peak vs.~the time to reach
that peak for all of the known RSNe.  The individual RSNe are labeled
and their types are indicated by their symbols; Type Ib, filled
squares, Type Ic, filled diamonds, and Type II filled triangles. The
lines of constant expansion velocity are shown for a brightness
temperature of $10^{11.5}$ K.}
\end{figure}

\subsubsection{Radio light curves}

A final, obvious comparison of SN 1998bw with other RSNe is the
evolution of its radio flux density at multiple frequencies.  The
radio data for the first 80 days are plotted in Kulkarni \etal\/(1998c),
up to day 250 on
{\it http://www.narrabri.atnf.csiro.au/public/grb\\/grb980425/}, and have been
kindly supplied to us in tabular form by D.~A.~Frail (private
communication).  SN 1998bw shows an early peak at higher frequencies
($\nu \ge 2.5$ GHz), which reaches a maximum as early as
day 10 -- 12 at 8.64 GHz, drops to a minimum almost simultaneously for the
higher frequencies at day $\sim$ 20 -- 24, and then reaches a secondary, somewhat
lower peak at later times after the first dip.  An interesting
characteristic of this ``double humped'' structure is that it dies out
at lower frequencies and is relatively inconspicuous in the 1.4 GHz
radio measurements (see Fig.~10).  Such a structure is not as
prominent for other known radio supernovae on such a short time scale,
although the Type Ic SN 1994I (S.~D.~Van Dyk, private communication)
shows a dip in its 15 GHz flux density, also at an age of $\sim20$
days.

\begin{figure}
\vspace{3cm}
\plotfiddle{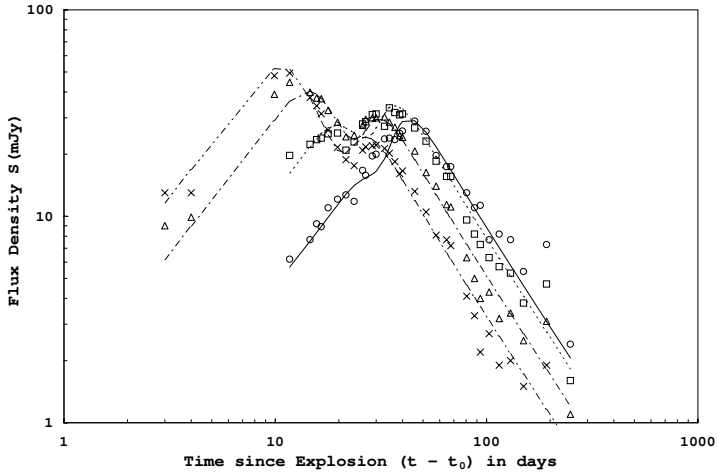}{4.0cm}{90}{40}{40}{160}{-20}
\caption{The radio light curves of SN 1998bw at 8.64 GHz (3.5 cm;
{\it cross}, {\it dash-double dot line}), 4.80 GHz (6.3 cm; {\it open triangle},
{\it dash-dot line}), 2.50 GHz (12 cm; {\it open square}, {\it dotted line}) and 1.38
GHz (21.7 cm; {\it open circle}, {\it solid line}).  The curves are derived from a
best fit model described by the equations in $\S$2.1 and the
parameters listed in Table 3.}
\end{figure}

Li \& Chevalier (2000) propose an initially synchrotron self-absorbed
(SSA), rapidly expanding shock wave in a $\rho \propto r^{-2}$
circumstellar wind model to describe the radio light curve.  This is
in many ways similar to the Chevalier (1998) model for Type Ib/c SNe, 
which also included
SSA, and the structure described in $\S$2.1, which includes only thermal
free-free initial absorption.  However, to improve the quality of the
fit, Li \& Chevalier (2000) introduce the additional free parameters
of a boost of shock energy by a factor of $\sim2.8$ on day $\sim$22 in the
observer's time frame.  This provides the energy necessary to produce
the second peak in the radio light curves.

While is not possible to rule out such complexities, we have attempted
to describe the radio light curve by a relatively straightforward,
two-component model, very similar to those used successfully for
previous SNe described above in $\S$2.1.  Such an approach has no requirement 
for shock re-acceleration.  Also, although an
attempt was made to include the effects of SSA at early times, the fit
was not improved by such an addition.  The two component model fit
which we obtained using only synchrotron emission and thermal, free-free absorption is shown as
the lines in Fig.~10, and its parameter values are listed in
Table 3.  The ``double humped'' structure of the radio light
curves for SN 1998bw is reproduced by a single energy shock wave encountering 
regions of differing
CSM density as it travels rapidly outward.

\begin{table}
\begin{center}
\begin{tabular}{ccc} 
{Parameter}         & {Component 1}          & {Component 2}           \\
$\alpha$            & -1.08                  &  -0.80                  \\ 
$\beta$             & -3.13                  &  -1.55                  \\ 
$K_1$               & $3.8 \times 10^5$      &  $6.2 \times 10^3$      \\  
$K_2$               & 0                      &  0                      \\
$\delta$            & --                      &  --                      \\
$K_3$               & $2.4 \times 10^5$      &  $8.4 \times 10^{12}$   \\
${\delta}^{\prime}$ & -4.43                  &  -8.80                  \\
$K_4$               & $4.3 \times 10^{-2}$   &  $3.4 \times 10^{-2}$   \\
\end{tabular}
\end{center}
\caption{Parameters for a Two Component, Thermal Model Fit (see $\S$2.1)} 
\end{table}

Since Li \& Chevalier (2000) do not give a quantitative
indication of the quality of their fit, it is impossible to compare the
two models directly.  However, a visual comparison of Fig.~10 here
with their Fig.~9 indicates that the fits are
of similar quality at late times and low frequencies, where the initial
absorption mechanisms have become less important.  This is not
surprising, since they are based on the same models from
Chevalier (1982a,b).  Both our model and the Li \& Chevalier (2000) model 
have difficulty adequately
describing the dip at all frequencies near day 20.  However, at early times and
high frequencies the free-free absorption model shown in Fig.~10 gives
a clearly superior fit to the data to that of Li \& Chevalier
(2000).  

Note that our fit requires essentially no diffuse/uniform absorption
($K_2$ = 0; see Table 3 and Eqs. 2.1 -- 2.4), so that all 
of the free-free absorption either is due to a clumpy
medium or is internal to the emitting zones (which has the same mathematical form),
 both of which contribute 
to a $K_3$ term in Table 3 and Eqs. 2.1 -- 2.4.  These results, combined
with the estimate of a high shock velocity, suggest that
the CSM around SN 1998bw is highly clumped in numerous and dense knots
with little, if any, intra-knot gas.  The clump filling factor has to be
high enough to intercept a considerable fraction of the shock energy and
still be low enough to let radiation escape from any given knot without
being appreciably absorbed by any other knot. This can be
somewhat quantified by postulating that the covering factor (\ie the fraction of
solid angle occupied by clumps) has to be less than $\sim$1/4 and, perhaps,
greater than $\sim$1/10. In this scenario the shock front can easily move at
a speed which is a significant fraction of the speed of light, because it is moving 
essentially in a vacuum.  However, one still has strong energy dissipation and
relativistic electron acceleration at the knot surfaces facing the SN
explosion center. If the knots are dense and relatively opaque to radio emission, 
we mainly observe radiation produced at the surfaces of the knots in the CSM 
on the far-side of the SN in
which some {\it internal} absorption is also occurring. The 
radiation from near-side knots is probably absorbed internally by
the knots themselves and, therefore, lost from the signal we
detect. 

The presence of two components, as evidenced by the presence of two
peaks in the radio light curve, implies the existence of two regimes
within the clumpy CSM that surrounds the SN progenitor with the transition
between the two regimes occurring at about 10 to 20 light-days from
the star, i.e., at about 3$\times$ 10$^{16}$ cm or $\sim$3000 AU from the
stellar progenitor. This distance may be somewhat speculatively interpreted as the separation
of the exploding primary from a lower-mass binary companion, which is
reminiscent of the binary structure inferred for SN 1979C (Weiler \etal\/1992a, 
Boffi \& Panagia 1996).  In the case of SN 1998bw a mass at explosion 
of about $\sim$14 M$_\odot$ and an original
progenitor mass of $\sim$30 -- 40 M$_\odot$ has been proposed (Danziger 
{\it et al.}, this Conference).  A binary hypothesis is also the best
explanation for the large mass loss that the progenitor star must have suffered
to expose the H-free layers typical of
Type Ib/c supernovae before exploding. Within this framework, one may argue that the CSM
within the SN progenitor Roche lobe was mostly determined by the
progenitor mass-loss itself, whereas at larger distances the CSM was
largely determined by the interaction of the progenitor wind with the 
companion wind, thus leading to a two-component CSM structure. 

Clearly, there is much more complexity in the physics of the GRB
phenomenon and the structure of the SN 1998bw system than can be described 
by simple models, and none proposed
is completely satisfactory as yet.  However, while the $\gamma$-ray
emission itself appears to require relativistic boosting, the radio
emission from the best-studied possible example, SN 1998bw, need not be
interpreted as proof of relativistic beaming, highly relativistic shock
waves, or other esoteric phenomena. Only mildly relativistic processes and
thermal absorption by the ionized CSM appear at least satisfactory and 
the arguments for a SN 1998bw/GRB 980425 connection are correspondingly weakened.

\subsubsection{Summary of SN 1998bw/GRB~980425 relations}

On balance, SN 1998bw appears to be a relatively normal, if rather
over-bright example of the SN Type Ib/c phenomenon.  Many observers
tend to accept that it is associated with GRB~980425, and it is
impossible to disprove that postulate.  On the other hand, neither the
optical nor radio emission are sufficiently unusual to firmly
establish a connection, and the Pian \etal\/(1999) X-ray source S2 in
the GRB field remains a distinct possibility for the location of
GRB~980425.  We can only hope that both the continued study of the GRB
phenomenon and its (sometimes) associated afterglows, and the slowly
increasing statistics of observations of Type Ib/c SNe, will be able to
establish the true nature of both phenomena and any possible relation.

\vskip .15in

\subsection{Acknowledgments}

CKL, KWW, \& MJM wish to thank the Office of Naval Research
(ONR) for the 6.1 funding supporting this research.


\smallskip

\begin{thebibliography}{}

\bibitem[Ball \etal\/1995]{Ball95}
{\sc Ball, L., Campbell-Wilson, D., Crawford, D. F.~\& Turtle, 
A. J.} 1995, \APJ {\bf 453} 864

\bibitem[Bartel \etal\/1985]{Bartel85}
{\sc Bartel, N.~\etal} 1985, \NAT {\bf 318} 25

\bibitem[Boffi \& Panagia 1996]{Boffi96}
{\sc Boffi, F. R.~\& Panagia, N.} 1996.  In {\it Radio Emission from 
the Stars and the Sun} (ed. A. R. Taylor \& J. M. Paredes).  Astr. Soc. 
Pacific Conf. Series, vol. 93, p. 153 

\bibitem[Bono \& Panagia 2000]{Bono00}
{\sc Bono, G.~\& Panagia, N.} 2000, in preparation.

\bibitem[Brocato \& Castellani 1993]{Brocato93}
{\sc Brocato, E.~\& Castellani, V.} 1993, \APJ {\bf 410} 99

\bibitem[Chevalier 1982a]{Chevalier82a}
{\sc Chevalier, R. A.} 1982a, \APJ {\bf 259} 302

\bibitem[Chevalier 1982b]{Chevalier82b}
{\sc Chevalier, R. A.} 1982b, \APJL {\bf 259} 85

\bibitem[Chevalier 1998]{Chevalier98}
{\sc Chevalier, R. A.} 1998, \APJ {\bf 499} 810

\bibitem[Chevalier \& Fransson 1994]{Chevalier94}
{\sc Chevalier, R. A.~\& Fransson, C.} 1994, \APJ {\bf 420} 268

\bibitem[Chu \etal\/1999]{Chu99}
{\sc Chu, Y.-H., Caulet, A., Montes, M. J., Panagia, N., Van Dyk,
S. D., \& Weiler, K.~W.} 1999, \APJL {\bf 512} 51.

\bibitem[Filippenko 1998]{Filippenko98}
{\sc Filippenko, A.} 1998, \IAUC 6969

\bibitem[Filippenko, Matheson, \& Barth 1994]{Filippenko94}
{\sc Filippenko, A., Matheson, T., \& Barth, A.} 1994, \AJ {\bf 108} 222

\bibitem[Frail \& Kulkarni 1997]{Frail97}
{\sc Frail, D. A.~\& Kulkarni, S. R.} 1997, \IAUC 6662

\bibitem[Fransson \& Bj\"orgsson 1998]{Fransson98}
{\sc Fransson, C.~\& Bj\"orgsson, C.-I.} 1998, \APJ {\bf 509} 861

\bibitem[Fruchter, \& Bergeron 1997]{Fruchter97}
{\sc Fruchter, A.~\& Bergeron, L.} 1997, \IAUC 6674

\bibitem[Gaensler \etal\/1997]{Gaensler97}
{\sc Gaensler, B. M., Manchester, R. N., Staveley-Smith, L.,
Tzioumis, A. K., Reynolds, J. E., \& Kesteven, M. J.} 1997, \APJ {\bf 479} 845

\bibitem[Galama \etal\/1998]{Galama98}
{\sc Galama, T. J.~\etal} 1998, \IAUC 6895

\bibitem[Garnavich, Kirshner, \& Challis 1997]{Garnavich97}
{\sc Garnavich, P., Kirshner, R., \& Challis, P.} 1997, \IAUC 6710

\bibitem[Green 1994]{Green94}
{\sc Green, D. W. E.} 1994, \IAUC 5953

\bibitem[Groot \etal\/1997]{Groot97}
{\sc Groot, P. J.~\etal} 1997, \IAUC 6584

\bibitem[Halpern \etal\/1997]{Halpern97}
{\sc Halpern, J.~\etal} 1997, \IAUC 6788

\bibitem[Hasinger, Aschenbach, \& Truemper 1996]{Hasinger96}
{\sc Hasinger, G., Aschenbach, B., \& Truemper, J.} 1996, \AAP {\bf 312} 9

\bibitem[Heise \etal\/1997]{Heise97}
{\sc Heise, J.~\etal} 1997, \IAUC 6787

\bibitem[Hoeflich \etal\/1996]{Hoeflich96}
{\sc Hoeflich, P., Wheeler, J. C., Hines, D. C., \& Tramaell,
S.~R.} 1996, \APJ {\bf 459} 307

\bibitem[Hamuy \etal\/1992]{Hamuy92}
{\sc Hamuy, M.~\etal} 1992, \IAUC 5574

\bibitem[Iwamoto \etal\/1998]{Iwamoto98}
{\sc Iwamoto, K.~\etal} 1998, \NAT {\bf 395} 672

\bibitem[Kay, Halpern, \& Leighly 1998]{Kay98}
{\sc Kay, L. E., Halpern, J. P., \& Leighly, K. M.} 1998, \IAUC 6969

\bibitem[Kellermann \& Pauliny-Toth 1969]{Kellermann69}
{\sc Kellermann, K. I.~\& Pauliny-Toth, I. I. K.} 1969, \APJL {\bf 155} 71

\bibitem[Kulkarni \etal\/1998a]{Kulkarni98a}
{\sc Kulkarni, S.~\etal} 1998a, \NAT {\bf 393} 35

\bibitem[Kulkarni \etal\/1998b]{Kulkarni98b}
{\sc Kulkarni, S. R., Bloom, J. S., Frail, D. A., Ekers, R.,
Wieringa, M., Wark, R., \& Higdon, J. L.} 1998b, \IAUC 6903

\bibitem[Kulkarni \etal\/1998c]{Kulkarni98c}
{\sc Kulkarni, S. R., Frail, D. A., Wieringa, M. H., Ekers, R. D.,
Sadler, E. M., Wark, R. M., Higdon, J. L., Phinney, E. S., \& Bloom,
J. S.} 1998c, \NAT {\bf 395} 663

\bibitem[Lacey \etal\/2000]{Lacey00}
{\sc Lacey, C. K., Weiler, K. W., Sramek, R. A., Panagia, N., \& Van
Dyk, S. D.} 2000, in preparation

\bibitem[Lang 1986]{Lang86}
{\sc Lang, K.R.} 1986. In {\it Astrophysical Formulae}, 47. Springer-Verlag

\bibitem[Langer \& Maeder 1995]{Langer95}
{\sc Langer, N.~\& Maeder, A.} 1995, \AAP {\bf 295} 685

\bibitem[Li \& Chevalier 2000]{Li00}
{\sc Li, Z.-Y.~\& Chevalier, R. A.} 2000, in press (also
astro-ph/9903483)

\bibitem[Lidman \etal\/1998]{Lidman98}
{\sc Lidman, C.~\etal} 1998, \IAUC 6895

\bibitem[Marcaide \etal\/1997]{Marcaide97}
{\sc Marcaide, J. M.~\etal} 1997, \APJL {\bf 486} 31

\bibitem[Miller, \& Branch 1990]{Miller90}
{\sc Miller, D. L.~\& Branch, D.} 1990, \AJ {\bf 100} 530

\bibitem[Montes, Weiler, \& Panagia 1997]{Montes97}
{\sc Montes, M. J., Weiler, K. W., \& Panagia, N.} 1997, \APJ {\bf 488} 792

\bibitem[Montes \etal\/1998]{Montes98}
{\sc Montes, M. J., Van Dyk, S. D., Weiler, K. W., Sramek, R. A., \&
Panagia, N.} 1998, \APJ {\bf 506} 874

\bibitem[Montes \etal\/2000]{Montes00}
{\sc Montes, M.~J., Weiler, K.~W., Van Dyk, S.~D., Sramek, R.~A.,
Panagia, N., \& Park, R.} 2000, {\it Astrophys.~J.}, in press (also
{\it astro-ph/9911399})
                                                             
\bibitem[Nomoto \etal\/1994]{Nomoto94}
{\sc Nomoto, K., Yamaoka, H., Pols, O. R., van den Heuvel, E.,
Iwamoto, K., Kumagai, S., \& Shigeyama, T.} 1994, \NAT {\bf 371} 227

\bibitem[Odewahn \etal\/1998]{Odewahn98}
{\sc Odewahn, S. C.~\etal} 1998, \APJ {\bf 509} 5

\bibitem[Paczynski 1998]{Paczynski98}
{\sc Paczynski, B.} 1998, \APJL {\bf 494} 45

\bibitem[Panagia \& Macchetto 1982]{Panagia82}
{\sc Panagia, N.~\& Macchetto, F.} 1982, \AAP {\bf 106} 266

\bibitem[Patat \etal\/1999]{Patat99}
{\sc Patat, F., Cappellaro, E., Rizzi, L., Turatto, M., \& Benetti,
S.} 1999, \IAUC 7215

\bibitem[Patat \& Piemonte 1998]{Patat98}
{\sc Patat, F.~\& Piemonte, A.} 1998, \IAUC 6918

\bibitem[Pian \etal\/1998]{Pian98}
{\sc Pian, E.~\etal} 1998, \APJL {\bf 492} 103

\bibitem[Pian \etal\/1999]{Pian99}
{\sc Pian, E.~\etal} 1999, \AAS {\bf 138} 463

\bibitem[Podsiadlowski \etal\/1993]{Podsiadlowski93}
{\sc Podsiadlowski, P., Hsu, J., Joss, P., \& Ross, R.} 1993, \NAT
{\bf 364} 509

\bibitem[Readhead 1994]{Readhead94}
{\sc Readhead, A. C. S.} 1994, \APJ {\bf 426} 51

\bibitem[Ryder \etal\/1993]{Ryder93}
{\sc Ryder, S., Staveley-Smith, L., Dopita, M., Petre, R., Colbert,
E., Malin, D., \& Schlegel, E.} 1993, \APJ {\bf 417} 167

\bibitem[Sadler \etal\/1998]{Sadler98}
{\sc Sadler, E. M., Stathakis, R. A., Boyle, B. J., \& Ekers,
R. D.} 1998, \IAUC 6901

\bibitem[Sahu \etal\/1997]{Sahu97}
{\sc Sahu, K. C.~\etal} 1997, \APJL {\bf 489} 127

\bibitem[Schwarz \& Pringle 1996]{Schwarz96}
{\sc Schwarz, D. H. \& Pringle, J. E.} 1996, \MN {\bf 282} 1018

\bibitem[Soffitta \etal\/1998]{Soffitta98}
{\sc Soffitta, P.~\etal} 1998, \IAUC 6884

\bibitem[Spyromilio 1994]{Spyromilio94}
{\sc Spyromilio, J.} 1994, \MN {\bf 266} 61

\bibitem[Tinney \etal\/1998]{Tinney98}
{\sc Tinney, C., Stathakis, R., Cannon, R., \& Galama, T.} 1998, \IAUC
6896

\bibitem[Tully 1988]{Tully88}
{\sc Tully, R. B.} 1988. In {\it Nearby Galaxies Catalogue}. Cambridge Univ.~Press.

\bibitem[Turtle \etal\/1987]{Turtle87}
{\sc  Turtle, A. J., Campbell-Wilson, D., Bunton, J. D., Jauncey,
D. L., Kesteven, M. J., Manchester, R. N., Norris, R. P., Storey,
M. C., \& Reynolds, J. E.} 1987, \NAT {\bf 327} 38

\bibitem[Van Dyk \etal\/1993]{Vandyk93}
{\sc Van Dyk, S. D., Sramek, R. A., Weiler, K. W., \& Panagia, N.} 1993, \APJL
{\bf 419} 69

\bibitem[Van Dyk \etal\/1994]{Vandyk94}
{\sc Van Dyk, S. D., Weiler, K. W., Sramek, R. A., Rupen, M., \& Panagia,
N.} 1994, \APJL {\bf 432} 115

\bibitem[Van Dyk, Hamuy, \& Filippenko 1996]{Vandyk96}
{\sc Van Dyk, S. D., Hamuy, M., \& Filippenko, A. V.} 1996, \AJ {\bf 111} 2017

\bibitem[Waxman \& Loeb 1999]{Waxman99}
{\sc Waxman, E.~\& Loeb, A.} 1999, \APJ {\bf 515} 721

\bibitem[Weiler \etal\/1986]{Weiler86}
{\sc Weiler, K. W., Sramek, R. A., Panagia, N., van der Hulst, J. M., \&
Salvati, M.} 1986, \APJ {\bf 301} 790

\bibitem[Weiler, Panagia, \& Sramek 1990]{Weiler90}
{\sc Weiler, K. W., Panagia, N., \& Sramek, R. A.} 1990, \APJ {\bf 364} 611

\bibitem[Weiler \etal\/1991]{Weiler91}
{\sc Weiler, K. W., Van Dyk, S. D., Panagia, N., Sramek, R. A., \& Discenna,
J.} 1991, \APJ {\bf 380} 161

\bibitem[Weiler \etal\/1992a]{Weiler92a}
{\sc Weiler, K. W., Van Dyk, S. D., Pringle, J., \& Panagia, N.} 1992a, \APJ
{\bf 399} 672

\bibitem[Weiler \etal\/1992b]{Weiler92b}
{\sc Weiler, K. W., Van Dyk, S. D., Panagia, N., \& Sramek, R. A.} 1992b, \APJ
{\bf 398} 248

\bibitem[Weiler \etal\/1998]{Weiler98}
{\sc Weiler, K. W., Van Dyk, S. D., Montes, M. J., Panagia, N., \&
Sramek, R. A.} 1998, \APJ {\bf 500} 51

\bibitem[Yin 1994]{Yin94}
{\sc Yin, Q. F.} 1994, \APJ {\bf 420} 152

\end{thebibliography}
\end{document}